\documentclass{article}
\usepackage{amsmath,amssymb}

\newcommand{\ri}{\mathrm{i}}

\newcommand{\rd}{\mathrm{d}}
\usepackage{graphicx}

\newcommand{\re}{\mathrm{Re}}
\begin{document}
\noindent{\LARGE\bfseries Externally driven one-dimensional Ising model}\\[2\baselineskip]
\textsf{Amir~Aghamohammadi}$^a$~\footnote{e-mail: mohamadi@alzahra.ac.ir},
\textsf{Cina~Aghamohammadi}$^b$~\footnote{e-mail: c\_aghamohammadi@yahoo.com}, \& \textsf{Mohammad~Khorrami}$^a$~\footnote{e-mail: mamwad@mailaps.org}\\[2\baselineskip]
$^a$Department of Physics, Alzahra University, Tehran 19938-91176, Iran\\
$^b$Department of Electrical Engineering, Sharif University of Technology, 11365-11155, Tehran, Iran


\begin{abstract}
\noindent A one dimensional kinetic Ising model at a finite temperature
on a semi-infinite lattice with time varying boundary spins is considered.
Exact expressions for the expectation values of the spin at
each site are obtained, in terms of the time dependent boundary
condition and the initial conditions. The solution consists of
a transient part which is due to the initial condition, and
a part driven by the boundary. The latter is an evanescent wave
when the boundary spin is oscillating harmonically. Low- and high-frequency
limits are investigated with greater detail. The total magnetization of
the lattice is also obtained. It is seen that for any arbitrary
rapidly varying boundary conditions, this total magnetization is equal to
the the boundary spin itself, plus essentially the time integral of
the boundary spin. A nonuniform model is also investigated.
\end{abstract}
\newpage
\section{Introduction}
Dynamical spin systems have played a central role in non-equilibrium
statistical models. The Ising model is widely studied in
statistical mechanics, as it is simple and allows one to understand
many features of phase transitions. The non-equilibrium properties of
the Ising model follow from the spin dynamics. In his article \cite{RG},
Glauber introduced a dynamical model formulating the dynamics
of spins, based on the rates coming from a detailed balance analysis.
It is a simple non-equilibrium model of interacting spins
with spin flip dynamics. An extension of the kinetic Ising model with
nonuniform coupling constants on a one-dimensional lattice was introduced
in \cite{DKSM}. In \cite{TV}, a damage spreading method was used to study
the sensitivity of the time evolution of a kinetic Ising model with
Glauber dynamics against the initial conditions. The full time dependence
of the space-dependent magnetization and of the equal time
spin-spin correlation functions were studied in \cite{SSG}.
Non-equilibrium two-time correlation and response functions for
the ferromagnetic Ising chain with Glauber dynamics have been studied
in \cite{GL,DRS}. The dynamics of a left-right asymmetric Ising
chain has been studied in \cite{God}. The response function to
an infinitesimal magnetic field for the Ising-Glauber model with
arbitrary exchange couplings was addressed in \cite{chat2003}.
In \cite{KA2002}, a Glauber model on a one-dimensional lattice
with boundaries was studied, for both
ferromagnetic and anti-ferromagnetic couplings.
The large-time behavior of the one-point function was studied.
It was shown that the system exhibits a dynamical phase transition,
which is controlled by the rate of spin flip at the boundaries.

It was shown in \cite{KA2008,KA2011}
that for a nonuniform extension of the kinetic Ising model,
there are cases where the system exhibits static and dynamical phase transitions.
Using a transfer matrix method, it was shown that there are
cases where the system exhibits a static phase transition,
which is a change of behavior of the static profile
of the expectation values of the spins near end points \cite{KA2008}.
Using the same method, it was shown in \cite{KA2011} that a dynamic
phase transition could occur as well: there is a fast phase
where the relaxation time is independent of the reaction rates
at the boundaries, and a slow phase where the relaxation time
does depend on the reaction rates at the boundaries.

Most of the studies on reaction diffusion models have
been on cases where the boundary conditions are constant
in time. Among the few models with time dependent
boundary conditions, is the asymmetric simple exclusion
process on a semi-infinite chain coupled at the end to
a reservoir with a particle density that changes periodically
in time \cite{PSS2008}. The situation is similar regarding
the case of the kinetic Ising model as well. Among the exceptions
are the study of the dynamical response of a two-dimensional
Ising model subject to a square-wave external field \cite{BR2008},
and the study of a harmonic oscillator linearly coupled with
a linear chain of Ising spins \cite{PBC2010,PBC2010-2}.

In this article a one dimensional Ising model at temperature $T$
on a semi-infinite lattice with time varying boundary spin is investigated.
The paper is organized as follows. In section 2 a brief review of
the formalism is presented, mainly to introduce the notation. In
section 3, a semi-infinite lattice with oscillating boundary spin is studied.
The exact solution for the expectation values of the spin at any site is obtained.
It is shown that there the boundary produces an evanescent wave in the lattice.
The low and high frequency limits are studied in greater detail.
The total magnetization of the lattice, $M(t)$, is also obtained.
It is shown that for rapidly changing boundary conditions,
the total magnetization is equal to the the boundary spin itself,
plus something proportional to the time integral of the boundary spin.
A nonuniform model in also investigated. It is shown that its evolution
operator eigenvalues are real. For the specific case of a
two-part lattice with each part being homogeneous, the reflection and
transmission coefficients corresponding to a harmonic source
at the end of the lattice are calculated. Finally, section 4 is
devoted to the concluding remarks.
\section{One-dimensional Ising model with nonuniform coupling constants}
Consider an Ising model on a one-dimensional lattice
with $L$ sites, labeled from $1$ to $L$. At each site
of the lattice there is a spin interacting with
its nearest neighboring sites according to the Ising Hamiltonian.
At the boundaries there are fixed magnetic fields. Denoting
the spin at the site $j$ by $s_j$, and the magnetic field at
the sites 1 and $L$ by $\mathfrak{B}_1$ and $\mathfrak{B}_L$,
one has for the Ising Hamiltonian
\begin{equation}\label{bo.1}
{\mathcal H} =-\sum_{\alpha=1+\mu}^{L-\mu}J_\alpha\,s_{\alpha-\mu}\,
s_{\alpha+\mu}-\mathfrak{B}_1\,s_1-\mathfrak{B}_L\,s_L.
\end{equation}
where $J_\alpha$ is the coupling constant in the link $\alpha$, and
\begin{equation}\label{bo.2}
\mu=\frac{1}{2}.
\end{equation}
The link $\alpha$ links the sites $\alpha-\mu$ and $\alpha+\mu$,
so that $\alpha\pm \mu$ are integers, and $\alpha$
runs from $\mu$ up to $(L-\mu)$. Throughout this paper, sites
are denoted by Latin letters which represent integers, while links
are denoted by Greek letters which represent integers plus one half
($\mu$). The spin variable $s_j$ takes the values $+1$ for spin up
($\uparrow$), or $-1$ for spin down ($\downarrow$). Define
\begin{equation}\label{bo.3}
K_\alpha:=\begin{cases}
\beta\,J_\alpha,& 1<\alpha<L\\
\beta\,\mathfrak{B}_1,& \alpha=\mu\\
\beta\,\mathfrak{B}_L,& \alpha=L+\mu
\end{cases}
\end{equation}
where
\begin{equation}\label{bo.4}
\beta:=\frac{1}{k_\mathrm{B}\,T},
\end{equation}
and $k_\mathrm{B}$ is the Boltzmann's constant, and
$T$ is the temperature. Denoting the reaction rate
from the configuration $A$ to the configuration
$B$ by $\omega(A\to B)$,
and assuming that in each step only one spin flips,
detailed balance demands the following for the reaction rates.
\begin{align}\label{bo.5}
\omega[(S', s_j)\to(S',-s_j)]&=\Gamma_j\,
[1-s_j\,\tanh(K_{j-\mu}\,s_{j-1}+K_{j+\mu}\,s_{j+1})],\nonumber\\
&\qquad 1<j<L,\\ \label{bo.6}
\omega[(S', s_1)\to(S',-s_1)]&=\Gamma_1\,
[1-s_1\,\tanh(K_\mu+K_{1+\mu}\,s_2)],\\ \label{bo.7}
\omega[(S', s_L)\to(S',-s_L)]&=\Gamma_L\,
[1-s_L\,\tanh(K_{L-\mu}\,s_{L-1}+K_{L+\mu})].
\end{align}
$\Gamma_j$'s are independent of the configurations. For simplicity,
we take them to be independent of the site. Then, rescaling
the time they are set equal to one.

So the evolution equation for the expectation value of the spin in
the site $j$ is
\begin{align}\label{bo.8}
\frac{\rd}{\rd t}\langle s_j\rangle&=-2\,\langle s_j\rangle+
[\tanh(K_{j-\mu}+K_{j+\mu})+\tanh(K_{j-\mu}-K_{j+\mu})]\,\langle
s_{j-1}\rangle\nonumber\\ &\quad+
[\tanh(K_{j-\mu}+K_{j+\mu})-\tanh(K_{j-\mu}-K_{j+\mu})]\, \langle
s_{j+1}\rangle, \quad 1<j<L,\nonumber\\
\frac{\rd}{\rd t}\langle s_1\rangle&=-2\,\langle s_1\rangle+
[\tanh(K_\mu+K_{1+\mu})+\tanh(K_\mu-K_{1+\mu})]\nonumber\\ &\quad+
[\tanh(K_\mu+K_{1+\mu})-\tanh(K_\mu-K_{1+\mu})]\,\langle s_2\rangle,\nonumber\\
\frac{\rd}{\rd t}\langle s_L\rangle&=-2\,\langle s_L\rangle+
[\tanh(K_{L-\mu}+K_{L+\mu})+\tanh(K_{L-\mu}-K_{L+\mu})]\,\langle
s_{L-1}\rangle\nonumber\\ &\quad+
[\tanh(K_{L-\mu}+K_{L+\mu})-\tanh(K_{L-\mu}-K_{L+\mu})].
\end{align}
These can be written in the form
\begin{align}\label{bo.9}
\frac{\rd}{\rd t}\langle s_j\rangle&=-2\,\langle s_j\rangle+
[\tanh(K_{j-\mu}+K_{j+\mu})+\tanh(K_{j-\mu}-K_{j+\mu})]\,\langle
s_{j-1}\rangle\nonumber\\ &\quad+
[\tanh(K_{j-\mu}+K_{j+\mu})-\tanh(K_{j-\mu}-K_{j+\mu})]\,\langle
s_{j+1}\rangle,\quad 1\leq j\leq L,\\ \label{bo.10}
\langle s_0\rangle&=1,\\ \label{bo.11}
\langle s_{L+1}\rangle&=1.
\end{align}
\section{Time varying boundary conditions on a semi-infinite lattice}
Consider a lattice for which the boundary spins ($s_0$ and
$s_{L+1}$) are externally controlled, but the reactions at
the internal sites satisfy detailed balance. The evolution
equation is then the same as (\ref{bo.9}), but combined
with boundary conditions different from (\ref{bo.10}) and
(\ref{bo.11}). A semi-infinite lattice the boundary of
which is externally controlled, is obtained by letting
$L$ tend to infinity, and using the following boundary
conditions
\begin{align}\label{bo.12}
&\langle s_0\rangle=f(t),\\ \label{bo.13}
&\mbox{$\langle s_j\rangle$ does not blow up as $j$ tends to infinity},
\end{align}
instead of (\ref{bo.10}) and (\ref{bo.11}).

A general solution of (\ref{bo.9}), combined with
(\ref{bo.12}) and (\ref{bo.13}), can be written
as the sum of a particular solution plus a general
solution of (\ref{bo.9}), combined with the homogeneous
boundary conditions.
\subsection{Semi-infinite lattice with uniform couplings: the homogeneous solution}
For a lattice with uniform couplings, $K_\alpha$'s are denoted by
$K$. The solution to the homogenous equation (vanishing $f$) is
denoted by $\langle s_j\rangle_\mathrm{h}$. One arrives at
\begin{equation}\label{bo.14}
\frac{\rd}{\rd t}\langle s_j\rangle_\mathrm{h}=-2\,\langle s_j\rangle_\mathrm{h}+
[\tanh(2\,K)]\,(\langle s_{j-1}\rangle_\mathrm{h}+\langle s_{j+1}\rangle_\mathrm{h}),\quad 0<j.
\end{equation}
Defining
\begin{equation}\label{bo.15}
\langle s_j\rangle_\mathrm{h}:=-\langle s_{-j}\rangle_\mathrm{h},\quad j<0,
\end{equation}
one arrives at
\begin{equation}\label{bo.16}
\frac{\rd}{\rd t}\langle s_j\rangle_\mathrm{h}=-2\,\langle s_j\rangle_\mathrm{h}+
[\tanh(2\,K)]\,(\langle s_{j-1}\rangle_\mathrm{h}+\langle s_{j+1}\rangle_\mathrm{h}),
\end{equation}
which holds for all integers $j$. Denoting  the linear  operator
acting on $\langle s_l\rangle$'s in the right-hand side of
(\ref{bo.16}) by $h$, the above equation is of the form
\begin{equation}\label{bo.17}
\frac{\rd}{\rd t}\langle s_j\rangle_\mathrm{h}=h^l_j\,\langle s_l\rangle_\mathrm{h},
\end{equation}
where $h^l_j$'s are the matrix elements of $h$.
Defining the generating function $G$ through
\begin{equation}\label{bo.18}
G(z,t):=\sum_{j=-\infty}^\infty\,z^j\,\langle s_j\rangle_\mathrm{h}(t),
\end{equation}
one arrives at
\begin{equation}\label{bo.19}
\frac{\partial G}{\partial t}=[-2+(z+z^{-1})\,\tanh(2\,K)]\,G,
\end{equation}
resulting in
\begin{align}\label{bo.20}
G(z,t)&=\exp\{[-2+(z+z^{-1})\,\tanh(2\,K)]\,t\}\,G(z,0),\nonumber\\
&=\exp(-2\,t)\,\sum_{k=-\infty}^\infty z^k\,\mathrm{I}_k[2\,t\,\tanh(2\,K)]\,G(z,0),\nonumber\\
&=\exp(-2\,t)\,\sum_{j=-\infty}^\infty z^j\,\sum_{l=-\infty}^\infty
\mathrm{I}_{j-l}[2\,t\,\tanh(2\,K)]\,\langle s_l\rangle_\mathrm{h}(0),
\end{align}
where $\mathrm{I}_k$ is the modified Bessel function of first kind of order $k$.
(\ref{bo.20}) results in
\begin{align}\label{bo.21}
\langle s_j\rangle_\mathrm{h}(t)&=\exp(-2\,t)\,\sum_{l=-\infty}^\infty
\mathrm{I}_{j-l}[2\,t\,\tanh(2\,K)]\,\langle s_l\rangle_\mathrm{h}(0),\nonumber\\
&=\exp(-2\,t)\,\sum_{l=1}^\infty\{\mathrm{I}_{j-l}[2\,t\,\tanh(2\,K)]-\mathrm{I}_{j+l}[2\,t\,\tanh(2\,K)]\}
\,\langle s_l\rangle_\mathrm{h}(0).
\end{align}
Using the large argument behavior of the modified Bessel functions,
it is seen that
\begin{equation}\label{bo.22}
\langle s_j\rangle_\mathrm{h}(t)\sim\exp\{-2\,[1-\tanh(2\,K)]\,t\},\quad j>0,
\end{equation}
showing that the homogeneous solution tends to zero at large times.
\subsection{Semi-infinite lattice with uniform couplings: the particular solution
corresponding to harmonic boundary conditions}
The harmonic boundary condition is
\begin{equation}\label{bo.23}
\langle s_0\rangle=\re[\sigma_0\,\exp(-\ri\,\omega\,t)]
\end{equation}
The following ansatz for a particular solution $\langle s_j\rangle_\mathrm{p}$
to equations (\ref{bo.9}) and (\ref{bo.12})
\begin{equation}\label{bo.24}
\langle s_j\rangle_\mathrm{p}=\re[\sigma_j\,\exp(-\ri\,\omega\,t)]
\end{equation}
results in
\begin{equation}\label{bo.25}
(\ri\,\omega-2)\,\sigma_j+[\tanh(2\,K)](\sigma_{j+1}+\sigma_{j-1})=0.
\end{equation}
This has a solution of the form
\begin{equation}\label{bo.26}
\sigma_j=c\,z^j,
\end{equation}
where $z$ satisfies
\begin{equation}\label{bo.27}
z+z^{-1}=\frac{-i\omega +2}{\tanh(2K)}.
\end{equation}
It is obvious that changing the sign of $K$ results
in changing the sign of $z$, while changing the sign of
$\omega$ results in changing $z$ to its complex conjugate.
So it is sufficient to consider only nonegative values of
$K$ and $\omega$. From now on, it is assumed that
$K$ and $\omega$ are nonnegative. (\ref{bo.27}) has
two solution for $z$, which are inverse of each other,
and none are unimodular. The boundary condition at infinity
imposes that of the two solutions of type (\ref{bo.26}),
only that solution is acceptable which corresponds to
the root of (\ref{bo.27}) with modulus less than
one. From now, only this root is denoted by $z$:
\begin{equation}\label{bo.28}
z:=r\,\exp(\ri\,\theta)
\end{equation}
where $r$ and $\theta$ are real and $r$ is positive
and less that one. The solution to (\ref{bo.25}) is then
\begin{equation}\label{bo.29}
\sigma_j=\sigma_0\,z^j.
\end{equation}
As $|z|$ is less than one, the particular solution (\ref{bo.24})
describes an evanescent wave. Obviously, the rate of decay
length and the phase speed, $\ell$ and $v$ respectively,
satisfy
\begin{align}\label{bo.30}
\ell&=-\frac{1}{\ln r},\nonumber\\
v&=\frac{\theta}{\omega}.
\end{align}
As the homogenous solution (\ref{bo.21}) tends to zero for large times,
the particular solution (\ref{bo.24}) is in fact the large times solution
to the problem of harmonic boundary condition.

Defining
\begin{align}\label{bo.31}
a&:=\frac{\omega}{2},\nonumber\\
b&:=\tanh 2\,K,\nonumber\\
u&:=\frac{r+r^{-1}}{2},
\end{align}
the real and imaginary parts of (\ref{bo.27}) read
\begin{align}\label{bo.32}
u\,\cos\theta&=\frac{1}{b},\nonumber\\
\sqrt{u^2-1}\,\sin\theta&=\frac{a}{b}.
\end{align}
So $u$ satisfies
\begin{equation}\label{bo.33}
b^2\,u^4-(a^2+b^2+1)\,u^2+1=0,
\end{equation}
from which one arrives, for the solution which is
larger than one, at
\begin{equation}\label{bo.34}
u=\left(\frac{1+a^2+b^2+\sqrt{(1+a^2+b^2)^2-4 b^2}}{2\,b^2}\right)^{1/2}.
\end{equation}
This is increasing with respect to $a$, and decreasing
with respect to $b$. Noting that
\begin{equation}\label{bo.35}
\frac{\rd u}{\rd r}=\frac{1}{2}\,\left(1-\frac{1}{r^2}\right),
\end{equation}
which shows that $u$ is decreasing with respect to $r$,
it is seen that $r$ is decreasing with respect to $\omega$,
and increasing with respect to $K$. One also has
\begin{equation}\label{bo.36}
r=u-\sqrt{u^2-1}.
\end{equation}

Regarding $\theta$, differentiating the first equation in
(\ref{bo.32})with respect to $a$, one has
\begin{equation}\label{bo.37}
\cos\theta\,\frac{\partial u}{\partial a}-u\,\sin\theta\,\frac{\partial\theta}{\partial a}=0,
\end{equation}
resulting in
\begin{align}\label{bo.38}
\frac{\partial\theta}{\partial a}&=\frac{\cos\theta}{u\,\sin\theta}\,
\frac{\partial u}{\partial a},\nonumber\\
&=\frac{b^2\,u^2-b^2}{b\,u\,\sqrt{(1+a^2+b^2)^2-4\,b^2}}\,\frac{\sin\theta}{a},\nonumber\\
&=\frac{b^2\,u^2-b^2}{b\,u\,(b^2\,u^2-u^{-2})}\,\frac{\sin\theta}{a}.
\end{align}
Equation (\ref{bo.34}) shows that
\begin{equation}\label{bo.39}
b\,u\geq 1,
\end{equation}
from which it is seen that
\begin{equation}\label{bo.40}
0<\frac{\partial\theta}{\partial a}\leq\frac{\sin\theta}{a}.
\end{equation}
The first inequality shows $\theta$ is an increasing function of $a$,
so it is an increasing function of $\omega$.
The second inequality results in
\begin{equation}\label{bo.41}
\frac{\partial\theta}{\partial a}\leq\frac{\theta}{a},
\end{equation}
which shows that $(\theta/a)$ is a decreasing function
of $a$. So $(a/\theta)$ is an increasing function of $a$, or
$(\omega/\theta)$ is an increasing function of $\omega$.

One also has
\begin{equation}\label{bo.42}
\frac{\partial(2\,b^2\,u^2)}{\partial b^2}=1+\frac{a^2+b^2-1}{\sqrt{(1+a^2+b^2)^2-4\,b^2}},
\end{equation}
and as
\begin{equation}\label{bo.43}
\sqrt{(1+a^2+b^2)^2-4\,b^2}\geq 1-b^2,
\end{equation}
it turns out that $(b\,u)$ is increasing with $b$, so that $\theta$
is increasing with $b$. Hence $(\omega/\theta)$ is decreasing with
$K$.

The asymptotic behavior of $r$ and $\theta$ is summarized as
\begin{equation}\label{bo.44}
r=\begin{cases}
\displaystyle{\tanh K},&
\omega\ll 1,\\ \\
\displaystyle{\frac{\tanh(2\,K)}{\omega}},& 1\ll\omega,\\ \\
\displaystyle{\frac{\tanh(2\,K)}{\sqrt{4+\omega^2}}},&
K\ll 1,\\ \\
\displaystyle{\frac{\sqrt{8+\omega^2+\omega\,\sqrt{16+\omega^2}}-
\sqrt{\omega^2+\omega\,\sqrt{16+\omega^2}}}{\sqrt{8}}},&1\ll K
\end{cases},
\end{equation}
and
\begin{equation}\label{bo.45}
\theta=\begin{cases}
\displaystyle{\frac{\omega\,\cosh(2\,K)}{2}},&
\omega\ll 1,\\ \\
\displaystyle{\frac{\pi}{2}},& 1\ll\omega,\\ \\
\displaystyle{\tan^{-1}\frac{\omega}{2}},&
K\ll 1,\\ \\
\displaystyle{\cos^{-1}\sqrt{\frac{8+\omega^2-\omega\,\sqrt{16+\omega^2}}{8}}},&1\ll K
\end{cases},
\end{equation}
Among other things, it is seen that the phase speed, at
low frequencies approaches the constant value
$2/[\cosh(2\,K)]$, while at high frequencies varies
like $(2\,\omega/\pi)$.

\begin{figure}
\begin{center}
\includegraphics[scale=0.3]{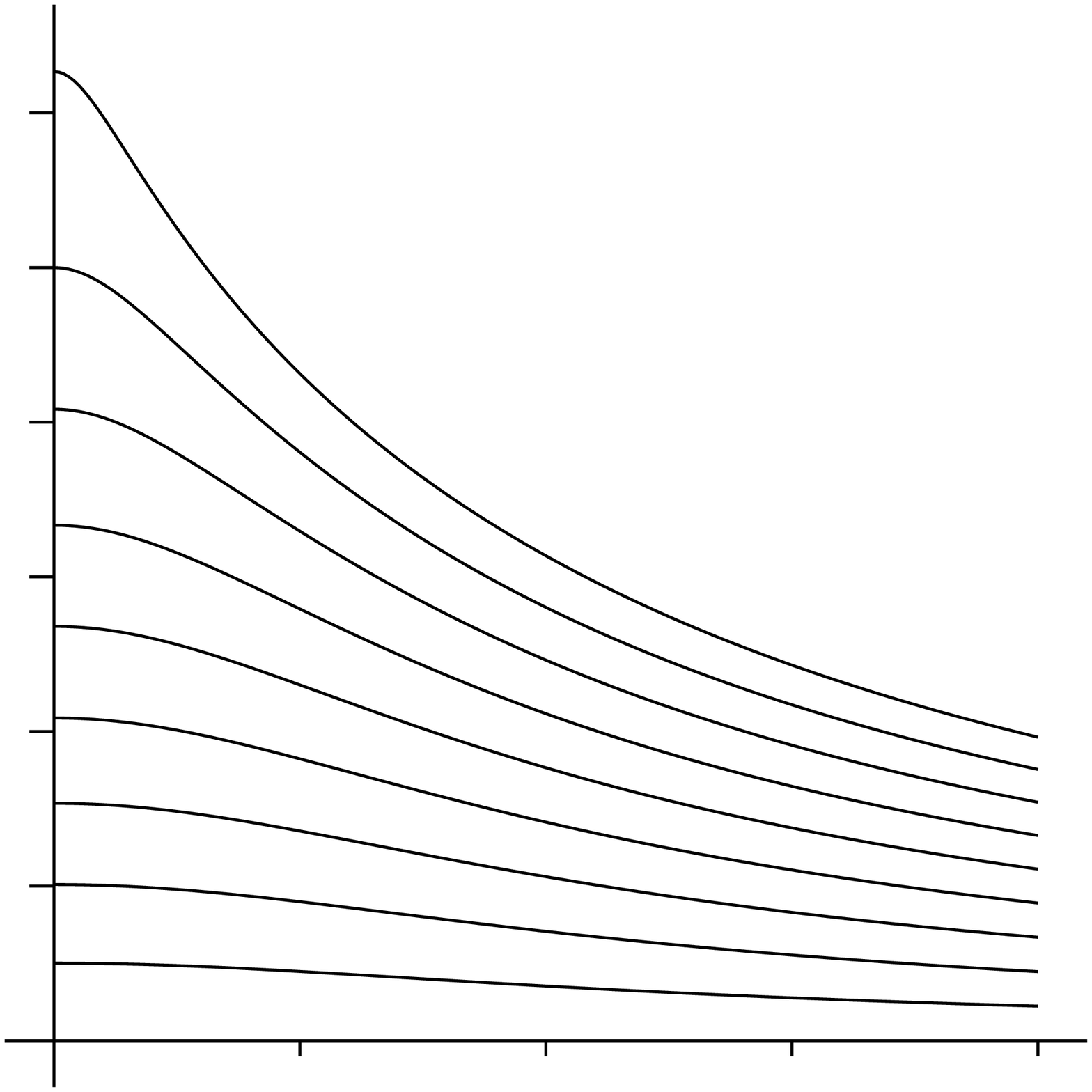}
\setlength{\unitlength}{1pt}
\put(-163,160){$r$}
\put(-7,4){$\omega$}
\put(-122,1){\scriptsize{1}}
\put(-172,31){\scriptsize{0.1}}
\put(-5,13){\scriptsize{0.1}}
\put(-5,53){\scriptsize{0.9}}
\put(-8,13){\vector(0,1){50}}
\put(-5,33){$\tanh(2\,K)$}
\caption{\label{fig1}}{The plot of $r$ versus $\omega$ for different values of $\tanh(2\,K)$}
\end{center}
\end{figure}

\begin{figure}
\begin{center}
\includegraphics[scale=0.3]{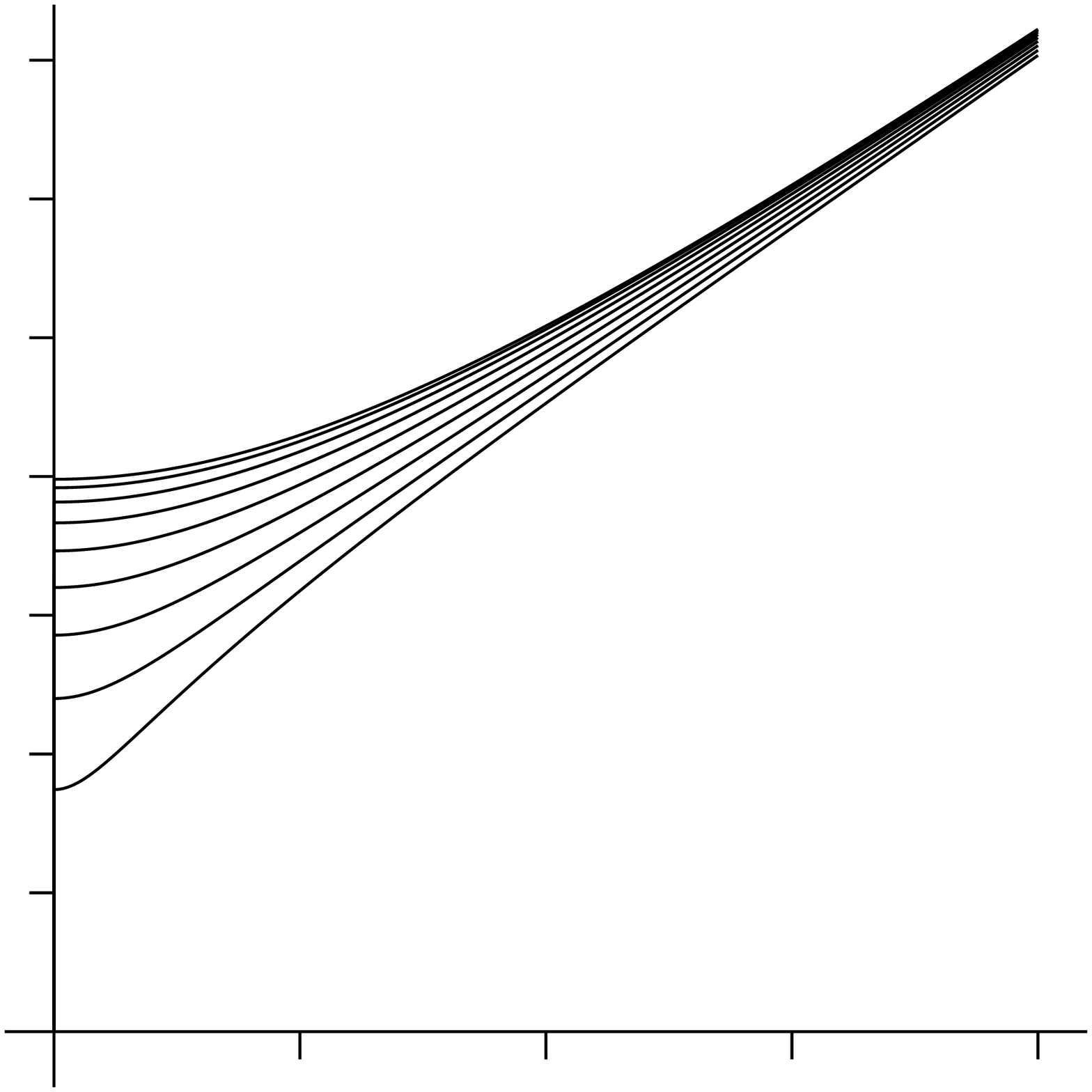}
\setlength{\unitlength}{1pt}
\put(-186,160){$(\omega/\theta)$}
\put(-7,4){$\omega$}
\put(-122,1){\scriptsize{1}}
\put(-166,51){\scriptsize{1}}
\put(-170,95){\vector(0,-1){55}}
\put(-182,92){\scriptsize{0.1}}
\put(-182,47){\scriptsize{0.9}}
\put(-216,69){$\tanh(2\,K)$}
\caption{\label{fig2}}{The plot of the phase speed $(\omega/\theta)$ versus $\omega$
for different values of $\tanh(2\,K)$}
\end{center}
\end{figure}

Figure \ref{fig1} is a plot of $r$ versus $\omega$ for different values of
$\tanh(2\,K)$ from $0.1$ to $0.9$. Figure \ref{fig2} is a plot of
the phase speed $(\omega/\theta)$ versus $\omega$ for
different values of $\tanh(2\,K)$ from $0.1$ to $0.9$.


The total magnetization, defined as the sum of the expectation
values of the spins, is denoted by $M$. At large times only
the particular solution contributes to the magnetization. So,
\begin{equation}\label{bo.46}
M=\re\left[\frac{\sigma_0}{1-z}\,\exp(-\ri\,\omega\,t)\right],\quad t\to\infty.
\end{equation}
For the time-independent boundary condition, this leads to
\begin{equation}\label{bo.47}
M=\frac{\sigma_0}{1-\tanh K},\quad (t\to\infty,\;\omega=0).
\end{equation}
For high frequencies,
\begin{equation}\label{bo.48}
M=\re\left\{\sigma_0\,\left[1-\frac{\tanh(2\,K)}{-\ri\,\omega}\right]^{-1}\,\exp(-\ri\,\omega\,t)\right\},
\quad (t\to\infty,\;\omega\to\infty).
\end{equation}
This can be simplified to
\begin{align}\label{bo.49}
M&=\re\left\{\sigma_0\,\left[1+\frac{\tanh(2\,K)}{-\ri\,\omega}\right]\,\exp(-\ri\,\omega\,t)\right\},
\nonumber\\ &=\langle s_0\rangle(t)+[\tanh(2\,K)]\,[S_0(t)-\bar S_0],
\quad (t\to\infty,\;\omega\to\infty),
\end{align}
where
\begin{align}\label{bo.50}
S_0(t)&:=\int_0^t\rd t'\,\langle s_0\rangle(t'),\nonumber\\
\bar S_0&:=\lim_{T\to\infty}\left[\frac{1}{T}\,\int_T\rd t\;S_0(t)\right].
\end{align}
One then arrives at a similar result for the magnetization
when the boundary condition is any arbitrary rapidly varying function
of time (so that its low frequency components are negligible):
\begin{align}\label{bo.51}
&M=\langle s_0\rangle(t)+[\tanh(2\,K)]\,[S_0(t)-\bar S_0],\nonumber\\
&\quad (t\to\infty,\mbox{ rapidly varying boundary conditions}).
\end{align}
\subsection{Semi-infinite lattice with two parts of uniform couplings: the particular solution
corresponding to harmonic boundary conditions}
Consider a semi-infinite lattice consisting of two parts, so that
\begin{equation}\label{bo.52}
K_\alpha=\begin{cases}K_1,&\alpha<N\\
K_2,&\alpha>N
\end{cases},
\end{equation}
The time evolution equations for the expectation values of the spins are
\begin{align}\label{bo.53}
\langle\dot s_j\rangle&=-2\,\langle s_j\rangle+
[\tanh(2\,K_1)]\,(\langle s_{j-1}\rangle+\langle s_{j+1}\rangle), \quad 0<j<N,\\ \label{bo.54}
\langle\dot s_N\rangle&=-2\,\langle s_N\rangle+\kappa_-\,\langle s_{N-1}\rangle+
\kappa_+\,\langle s_{N+1}\rangle,\\ \label{bo.55}
\langle\dot s_j\rangle&=-2\,\langle s_j\rangle+
[\tanh(2\,K_2)]\,(\langle s_{j-1}\rangle+\langle s_{j+1}\rangle), \quad N<j.
\end{align}
where
\begin{align}\label{bo.56}
\kappa_-&:=\tanh(K_1+K_2)+\tanh(K_1-K_2),\nonumber\\
\kappa_+&:=\tanh(K_1+K_2)-\tanh(K_1-K_2).
\end{align}
Applying a harmonic boundary condition (\ref{bo.23}), one has for
the particular solution of the kind (\ref{bo.24}),
\begin{equation}\label{bo.57}
\sigma_j=\begin{cases}(A_1\,z_1^j+B_1\,z_1^{-j}),& 0\leq j\leq N\\
A_2\,z_2^j,& N\leq j\end{cases},
\end{equation}
where
\begin{equation}\label{bo.58}
z_l+ z_l^{-1}=\frac{-\ri\,\omega+2}{\tanh(2\,K_l)},\quad l=1,2
\end{equation}
and $|z_l|$ is smaller than one. The boundary condition results in
\begin{equation}\label{bo.59}
A_1+B_1=\sigma_0.
\end{equation}
From (\ref{bo.57}) for $j=N$, one arrives at
\begin{equation}\label{bo.60}
A_1\,z_1^N+B_1\,z_1^{-N}=A_2\,z_2^N.
\end{equation}
Finally, (\ref{bo.54}) results in
\begin{equation}\label{bo.61}
\kappa_-\,(A_1\,z_1^{N-1}+B_1\,z_1^{-N+1})+\kappa_+\,A_2\,z_2^{N+1}=(-\ri\,\omega +2)\,A_2\,z_2^N.
\end{equation}
Equations (\ref{bo.59}) through (\ref{bo.61}) give
\begin{align}\label{bo.62}
A_1&=\frac{(\kappa_-\,z_1+\kappa_+\,z_2+\ri\,\omega-2)\,z_1^{-N}\,\sigma_0}
{\kappa_-\,(z_1^{-N+1}-z_1^{N-1})+(\kappa_+\,z_2+\ri\,\omega-2)\,(z_1^{-N}-z_1^N)},\nonumber\\
B_1&=\frac{-(\kappa_-\,z_1^{-1}+\kappa_+\,z_2+\ri\,\omega-2)\,z_1^N\,\sigma_0}
{\kappa_-\,(z_1^{-N+1}-z_1^{N-1})+(\kappa_+\,z_2+\ri\,\omega-2)\,(z_1^{-N}-z_1^N)},\nonumber\\
A_2&=\frac{\kappa_-\,(z_1-z_1^{-1})\,z_2^{-N}\,\sigma_0}
{\kappa_-\,(z_1^{-N+1}-z_1^{N-1})+(\kappa_+\,z_2+\ri\,\omega-2)\,(z_1^{-N}-z_1^N)}.
\end{align}
For large $N$, these simplify to
\begin{align}\label{bo.63}
A_1&=\sigma_0,\nonumber\\
B_1&=\frac{-(\kappa_-\,z_1^{-1}+\kappa_+\,z_2+\ri\,\omega-2)\,z_1^{2\,N}\,\sigma_0}
{\kappa_-\,z_1+\kappa_+\,z_2+\ri\,\omega-2},\nonumber\\
A_2&=\frac{\kappa_-\,(z_1-z_1^{-1})\,z_2^{-N}\,z_1^N\,\sigma_0}
{\kappa_-\,z_1+\kappa_+\,z_2+\ri\,\omega-2}.
\end{align}
It can be easily shown that for the nonuniform lattice and at
high frequencies, up to first term in $\omega^{-1}$ the magnetization
is similar to the case of the uniform lattice.
\subsection{Semi-infinite lattice with nonuniform couplings:
the relaxation times}
The general solution of (\ref{bo.9}) with (\ref{bo.12}) and (\ref{bo.13})
is the sum of a particular solution and the general solution to
(\ref{bo.9}) and (\ref{bo.12}) and (\ref{bo.13}) with vanishing $f$.
The latter (the homogeneous solution) satisfies
\begin{align}\label{bo.64}
\frac{\rd}{\rd t}\langle s_j\rangle_\mathrm{h}&=-2\,\langle s_j\rangle_\mathrm{h}+
[\tanh(K_{j-\mu}+K_{j+\mu})+\tanh(K_{j-\mu}-K_{j+\mu})]\,\langle
s_{j-1}\rangle_\mathrm{h}\nonumber\\ &\quad+
[\tanh(K_{j-\mu}+K_{j+\mu})-\tanh(K_{j-\mu}-K_{j+\mu})]\,\langle
s_{j+1}\rangle_\mathrm{h},\quad 1\leq j\leq L,\nonumber\\
\langle s_0\rangle_\mathrm{h}&=0,\nonumber\\
\langle s_{L+1}\rangle_\mathrm{h}&=0,
\end{align}
which can be written as (\ref{bo.17}). Denoting an eigenvalue
of $h$ by $E$, and the corresponding eigenvector by $\psi_E$,
it is seen that there are solutions to (\ref{bo.64}) of the form
\begin{equation}\label{bo.65}
\langle s_j\rangle_\mathrm{h}(t)=\psi_{E\,j}\,\exp(E\,t).
\end{equation}
These solutions decay with a relaxation time $\tau$ satisfying
\begin{equation}\label{bo.66}
\tau=-\frac{1}{\mathrm{Re}(E)}.
\end{equation}

One can see that the eigenvalues of the operator $h$ are real.
To see this, one notices that equations (\ref{bo.64}) are
the same as the equations corresponding to the homogeneous
solution of (\ref{bo.8}). So the homogeneous solution to
the Ising chain externally driven at ends, is the same as
the homogenous solution to the Ising chain with magnetic
fields at boundaries. The evolution equation for the latter
satisfies the detailed balance. For any evolution satisfying
detailed balance, the eigenvalues of the evolution operator
are real. To see this, one notices that the criterion of
the detailed balance is
\begin{equation}\label{bo.67}
\omega(B\to A)=Y^A_B\,\exp[\beta(\mathcal{E}_B-\mathcal{E}_A)],
\end{equation}
where $A$ and $B$ are two different state, $Y^A_B$'s are real nonnegative
numbers (for $B\ne A$) satisfying
\begin{equation}\label{bo.68}
Y^B_A=Y^A_B,
\end{equation}
and $\mathcal{E}$ is the energy of the system in the state $A$.
So the matrix $Y$ is Hermitian. Equation (\ref{bo.67})
means that the evolution matrix $H$, the components of
which are $\omega(B\to A)$'s, is a similarity-transformed of
$Y$. As $Y$ is Hermitian, the eigenvalues of $Y$ are real.
As $H$ is a similarity-transformed of $Y$, the eigenvalues of
$H$ are the same as the eigenvalues of $Y$. So
the eigenvalues of $H$ are real (\cite{db} for example).
The eigenvalues of $h$ are eigenvalues of $H$ as well. So
the eigenvalues of $h$ are real.
\section{Concluding remarks}
A one dimensional kinetic Ising model at temperature $T$,
with time varying boundary conditions was studied, for the case
the lattice is semi-infinite. The evolution equation for
the expectation values of the spins was investigated. For
the case of harmonic boundary conditions, with uniform couplings,
exact particular solutions were obtained for the expectation values
of the spins, as well as the total magnetization. The low- and
high-frequency behaviors were studied in more detail.
Models for which the coupling constant is nonuniform were
also studied. Physically, such nonuniform couplings
could arise when either the interaction between spins or
the temperature depends on the position.
As a specific example, the harmonic solution on
a semi-infinite lattice consisting of two homogeneous
parts studied.
Finally, it was shown that for a general (nonuniform) lattice,
the eigenvalues corresponding to the evolution operator are real.
\\[\baselineskip]
\textbf{Acknowledgement}:  This work was supported by
the research council of the Alzahra University.

\end{document}